\documentclass[aps,prb, twocolumn, superscriptaddress]{revtex4-1}
\usepackage{graphicx}
\usepackage{amsmath}
\usepackage{amsfonts}
\usepackage{amsthm}
\usepackage{amssymb}
\usepackage{amsbsy}
\usepackage{pstricks}
\usepackage{wasysym}
\usepackage{bm}
\usepackage{mathrsfs}

\newcommand{\beq}{\begin{equation}}
\newcommand{\eeq}{\end{equation}}
\newcommand{\bea}{\begin{eqnarray}}
\newcommand{\eea}{\end{eqnarray}}

\def\xhat{\hat{\mathbf{x}}}
\def\yhat{\hat{\mathbf{y}}}
\def\zhat{\hat{\mathbf{z}}}

\def\ket#1{{\left|#1\right\rangle}}
\def\bra#1{{\left\langle #1 \right|}}

\def\br{\mathbf{r}}
\def\bk{\mathbf{k}}
\def\bq{\mathbf{q}}
\def\bp{\mathbf{p}}
\def\bR{\mathbf{R}}

\def\Winf{W_\infty}

\def\be{\begin{eqnarray}}
\def\ee{\end{eqnarray}}

\begin{document}

\title{Fractional Chern Insulators and the $W_\infty$ Algebra}
\author{S.  A. Parameswaran}\altaffiliation{Current address: Department of Physics, University of California, Berkeley, California 94720, USA}
\email{sidp@berkeley.edu}
\affiliation{Department of Physics, Princeton
University, Princeton, NJ 08544, USA}
\author{R. Roy}\altaffiliation{Current address: Department of Physics and Astronomy, University
of California, Los Angeles, California 90095-1547,
USA}
\email{rroy@physics.ucla.edu}Ê
\affiliation{Rudolf Peierls Centre for Theoretical Physics, Oxford University, 1 Keble Road, Oxford OX1 3NP, United Kingdom}
\author{S. L. Sondhi}
\email{sondhi@princeton.edu}
\affiliation{Department of Physics, Princeton
University, Princeton, NJ 08544, USA}

\date{\today}
\begin{abstract}
A set of recent results indicates that fractionally filled bands of Chern insulators in
two dimensions support fractional quantum Hall states analogous to those found in fractionally
filled Landau levels. We provide an understanding of these results by examining the algebra of
Chern band projected density operators. We find that this algebra closes at long wavelengths
and for constant Berry curvature, whereupon it is isomorphic to the $W_\infty$ algebra of lowest
Landau level projected densities first identified by Girvin, MacDonald and Platzman [Phys. Rev. B {\bf 33}, 2481 (1986).] For
Hamiltonians projected to the Chern band this provides a route to replicating lowest Landau
level physics on the lattice. 
\end{abstract}
\maketitle
\noindent

\noindent{\bf Introduction:}
Much of the theory of the quantum Hall effect is tied to the celebrated
Landau levels (LLs)---the non-dispersing energy states of free
electrons in a uniform magnetic field.  A fruitful line of inquiry has
focused on separating quantum Hall physics from this weak lattice limit.
In the case of the integer quantum Hall effect (IQHE), Thouless and co-workers \cite{TKNN}
showed that the Hall conductance is quantized even in the
presence of a strong periodic potential, where the energy bands are no
longer flat, as long as the chemical potential lies in a gap
between the bands. They did so by relating the conductance to a mathematical invariant,
the first Chern number, associated with the filled bands. Subsequently,
Haldane \cite{Haldane:1988p1} showed, through an explicit construction of a tight-binding model of a ``Chern
band insulator'' (CBI) that a quantized Hall conductance could be obtained even in the
absence of a net magnetic field so long as the filled bands still came with
non-trivial topology i.e., nonzero Chern numbers. This thus established an equivalence
between filled CBs and filled LLs.  A natural question that
then arises is whether such an equivalence also holds for partially filled CBs
and partially filled LLs---specifically whether the fractional quantum Hall
effect (FQHE) arises in a partially filled CB in a lattice system.

A flurry of recent work has shed much light on this question.
Several authors \cite{Tang:2011p1, Sun:2011p1, Neupert:2011p1, Hu:2011p1} have constructed models with nearly flat (non-dispersing) Chern
bands which allow interactions to dominate at partial fillings while
leaving the gap to neighboring bands open. Refs. \cite{Neupert:2011p1,  Sheng:2011p1, Wang:2011p1}   and, most
convincingly, Regnault and
Bernevig \cite{Regnault:2011p1} have reported evidence for FQH states at $\nu=1/3$, $1/5$ and (for bosons) $1/2$,  in finite-size studies of short-ranged interactions projected to
these bands. 
A partial understanding of these numerical results is already
available: applying a generalized Pauli principle familiar
from the lowest Landau level (LLL) FQHE  rationalizes some aspects of the results \cite{Regnault:2011p1}, while Qi \cite{Qi:2011p1} has constructed
a fairly general recipe for translating familiar model wavefunctions and Hamiltonians
from the LLL to CBs on cylinders by elegantly mapping Landau gauge eigenfunctions to particular
Wannier functions.

In this Letter, we offer a complementary perspective on the CB---LLL equivalence. As
in the earlier work we start by distinguishing the kinematical and dynamical aspects of
obtaining QH physics in a CB. The latter involve an appropriate choice of band structure
and interaction Hamiltonian but the former requires an appropriate identification of the
algebra of operators in the two systems. Specifically, in both the CB and the LLL,
the projected Hamiltonians that include interactions and scalar disorder involve 
only the projected particle density operators and thus the nontriviality of the physics enters precisely through the nontriviality of the algebraic relations between the different Fourier
components of the densities. For the LLL the density operators obey the so-called $W_\infty$ algebra
first constructed by Girvin, Macdonald and Platzman \cite{GMP:1986p1}. We show that at long wavelengths and in a natural limit (constant Berry
curvature) densities projected to the CB obey the {\it same} algebra. This implies that
{\it mutatis mutandis} LLL physics can be transported to the CB. We also note that this implies
that the cartesian components of the particle positions projected to the CB fail to
commute, much like guiding center coordinates in the LLL. In what follows we review the $\Winf$
algebra for the LLL, show how it arises in a CB  in $d=2$, argue that it allows LLL physics
(impurity and FQHE) to be translated to the CB and discuss
generalization of the density algebra to other topological insulators, which appears to us to
be a promising route to understanding their fractional physics.

\noindent{\bf Projection to the Lowest Landau Level:} We begin by recalling some details of
the projection to the LLL \footnote{Similar arguments apply to {\it any} LL as does the equivalence to the CB discussed in the following.} which captures the essential features of the
FQHE in the high field limit \cite{Shankar:Cargese2002}. Consider an electron in an external magnetic field $\mathbf{B} = -B \zhat$, described by the Hamiltonian $H=\frac{1}{2m}\left(\mathbf{p} + \frac{e}{c} \mathbf{A}\right)^2$. The motion of the electron can be separated into the fast cyclotron motion and the slower drift of the guiding center of its
orbit. Working in symmetric gauge and with $\hbar=1$, this can be implemented by decomposing its position $\br$
into two parts, $\br =  (\frac{1}{2}\br + \zhat\times\bp) +   (\frac{1}{2}\br - \zhat\times\bp)  \equiv {\bm \eta} + \bR$ which obey the commutation relations $[\eta_x, \eta_y] = i\ell_B^2 ,
[R_x, R_y] = -i\ell_B^2, [\eta_i, R_j] = 0$ where $\ell_B = (c/eB)^{1/2} $ is the magnetic length. In terms of these variables, the Hamiltonian depends purely on the cyclotron
coordinates $\bm\eta$,
$H = \frac{\bm \eta^2}{2m \ell_B^4}$, which (owing to the fact that the components of $\bm\eta$ are conjugate) has the familiar oscillator spectrum of a series of Landau levels, with
$E_n = \left(n +\frac{1}{2}\right) \omega_c$ where $\omega_c = eB/mc$ is the cyclotron frequency. Since $H$ commutes with the guiding center coordinates $\bR$, each energy level is extensively degenerate with a degeneracy  given by $N_\Phi$, the number of flux quanta threading the sample.

When the number of electrons $N$ is smaller than $N_\Phi$, we have a partially filled $n=0$ LL massively degenerate in the  kinetic energy; we must therefore include the effect of interactions. A reasonable starting point is to project the interactions into the degenerate subspace;
the effect of the projection is to replace $\bm\eta$ -dependent  expressions by their LLL expectation values e.g. $\langle\bm\eta^2\rangle_{LLL} = \ell_B^2$. The density operator, $\rho_{\bq} =e^{i\bq\cdot\hat{\br}}$ is projected as 
$\mathcal{P} \rho_\bq \mathcal{P}  =  e^{-q^2\ell_B^2/4} e^{i\bq\cdot\bR} \equiv e^{-q^2\ell_B^2/4} \overline{\rho}_{\bq}
$
where we have factored out a $q$-dependent constant in the definition of the projected density \footnote{As defined, $\overline{\rho}$ is  the magnetic translation operator, but in the LLL the distinction is unimportant for us.}. Since $[R_x, R_y] = -i\ell_B^2$, we can show that the $\overline{\rho}_\bq$ satisfy the 
$\Winf$/Girvin-MacDonald-Platzman algebra \cite{GMP:1986p1}
\be\label{eq:GMP}
[\overline{\rho}_{\bq_1},\overline{\rho}_{\bq_2}] =  2i \sin \left(\frac{\bq_1\wedge\bq_2\ell_B^2}{2}\right) \overline{\rho}_{\bq_1+\bq_2}
\ee
where $\bq_1\wedge\bq_2 \equiv \zhat\cdot(\bq_1\times\bq_2)$.  This algebra arises in several contexts and has the interpretation of a quantum deformation of the algebra of area-preserving diffeomorphisms on the plane as well as that of magnetic translations in a uniform field as discussed, e.g. in \cite{koganreview}.

Along with the above commutation relations, the Hamiltonian
\be\label{eq:LLLHamiltonian}
H_{LLL} = \frac{1}{2} \sum_{\bq} V(\bq)e^{-q^2\ell_B^2/2} \overline{\rho}_\bq \overline{\rho}_{-\bq} \nonumber \\
+ \sum_{\bq} V_{\rm imp} (\bq)e^{-q^2\ell_B^2/4} \overline{\rho}_{-\bq} \ ,
\ee
which is the projection of a density-density interaction and an impurity potential into the LLL, completes the low-energy description of the fractional quantum Hall effect. [In passing
from (\ref{eq:GMP}) to (\ref{eq:LLLHamiltonian}) we reinterpret $\overline{\rho}_\bq$ as the density of a many electron system. As the latter is additive over the individual particle
densities, it also obeys the algebra (\ref{eq:GMP}).]
While still formidable, the LLL problem has been tackled by a variety of different approaches, most notably by guessing trial wavefunctions for FQH states in the clean limit. The essential point for this Letter is that if we
can construct a similar low-energy description for a Chern insulator, then some aspects\footnote{While trial wave functions, model Hamiltonians, and so on may be mapped between the two problems, some aspects specific to Landau level physics -- such as the correspondence with conformal field theory correlators -- remain a special property of the FQHE in an external field; we thank an anonymous referee for reminding us of this.} of the familiar fractional quantum Hall technology can be applied -- with suitable modifications to account for the different nature of the single-particle states  -- to the problem of interacting Chern insulators.

\noindent{\bf Projection to the Chern Band:}
We begin by fixing notation and recalling some basic facts. An insulator with  $\mathcal{N}$ bands is described by the single-particle Hamiltonian
$H = \sum_{\bk,a,b} c^\dagger_{\bk,a} h_{ab}(\bk) c_{\bk,b}$
where $a,b =1,2,\ldots,\mathcal{N}$ are sublattice/spin indices, and $\bk$ is the crystal momentum restricted to the first Brillouin zone (BZ). (Here and below, we will explicitly indicate  when repeated indices are summed over.)
The solution of the $\mathcal{N}\times\mathcal{N}$ eigenvalue problem
$\sum_{b} h_{ab}(\bk) u^{\alpha}_b(\bk) = \epsilon_{\alpha}(\bk) u^{\alpha}_a(\bk)$
defines the energy bands and we will take the eigenvectors to be normalized, $\sum_{a} |u^{\alpha}_{a}(\bk)|^2 =1$.
The corresponding eigenstates are given by
\be\label{eq:eigstate}
\ket{\bk,\alpha} = \gamma^\dagger_{\bk,\alpha}\ket{0} &\equiv& \sum_a u^{\alpha}_{a}(\bk)c^\dagger_{\bk,a} \ket{0}
\ee
Specializing to two dimensions, the Chern number of a given band $\alpha$ is computed as
\be\label{eq:Chernnumber}
C_\alpha = \frac{1}{2\pi} \int _{\text{BZ}}{d^2 k}\, \mathcal{B}_\alpha(\bk).\ee
Here, $\mathcal{B}_\alpha(\bk)$ is the  Chern flux density (Berry curvature), defined as the curl of the Berry connection (Berry gauge potential), $\mathcal{B}_\alpha(\bk) = \nabla_{\bk} \times \mathcal{A}_\alpha(\bk)$. In terms of the eigenstates, we have 
\be\label{eq:Chernconnection}
 \mathcal{A}_\alpha(\bk) =  i \sum_{b=1}^\mathcal{N} u^{\alpha*}_b(\bk) \nabla_{\bk} u^\alpha_b(\bk) .
\ee
A filled band with Chern number $C_\alpha$ yields a Hall conductance $\sigma_H = C_\alpha e^2/h$ regardless of whether it arises in a system with a net magnetic field \cite{TKNN}
(``Hofstadter band'') or zero net magnetic field \cite{Haldane:1988p1} (``Haldane band''). We shall refer to both as Chern bands.

As announced, we are interested in the restriction of the kinematics to a single CB which can be implemented by use of the projection operator
$\mathcal{P_\alpha} = \ket{\bk,\alpha}\bra{\bk,\alpha}$. 
It follows that the density operator $\rho_\bq = e^{i\bq\cdot\hat{\br}}$ when projected onto the CB takes the form
\be\label{eq:projecteddensity}
\overline{\rho}_\bq \equiv \mathcal{P_\alpha}\rho(\bq) \mathcal{P_\alpha} &=& \sum_{\bk}
\left( \sum_b u^{\alpha *}_b\left(\bk+\frac{\bq}{2}\right) u^\alpha_b \left(\bk-\frac{\bq}{2}\right) \right) \nonumber \\
& &\times \gamma^\dagger_{\bk+\frac{\bq}{2},\alpha}\gamma_{ \bk-\frac{\bq}{2},\alpha}
\ee
At long wavelengths $qa\ll1$, we may expand $\sum_{b}  u^{\alpha *}_b\left(\bk+\frac{\bq}{2}\right)u^{\alpha}_{b}\left(\bk-\frac{\bq}{2}\right)
\approx 1 - i\bq \cdot \sum_b u_b^{\alpha *}(\bk) \frac{\nabla_\bk}{i} u^\alpha_b (\bk)$
$ \approx e^{i \int_{\bk-\bq/2}^{\bk+\bq/2} d\bk' \cdot\mathcal{A_\alpha}(\bk')}$, so that
\be\label{eq:approxparalleltransport}
\overline{\rho}_\bq \ket{\bk, \alpha} \approx e^{i \int_{\bk}^{\bk+\bq} d\bk' \cdot\mathcal{A_\alpha}(\bk')} \ket{\bk +\bq, \alpha} \ .
\ee
 In other words, for small $q$, $\overline{\rho}(\bq)$ implements parallel transport described by the Berry connection 
  $\mathcal{A}_\alpha(\bk)$. 
Either from this observation or via a gradient expansion, we may show that at long wavelengths, the commutator of projected density operators at different wavevectors is
\be\label{eq:Cherncomm1}
\left[ \overline{\rho}_{\bq_1},\overline{\rho}_{\bq_2}\right] &\approx& i \, \bq_1\wedge\bq_2 \sum_{\bk} \left[\mathcal{B}_\alpha(\bk)\sum_b   u^{\alpha*}_b\left(\bk_+\right)u^\alpha_b\left(\bk_-\right)\right.\nonumber\\& & \,\,\,\,\,\,\,\,\,\,\,\,\,\,\,\,\,\,\,\,\,\,\,\,\,\,\,\,\left.\times \gamma^\dagger_{\bk_+,\alpha}\gamma_{\bk_-,\alpha}\right]
\ee
where we define $\bk_{\pm} = \bk \pm \frac{\bq_1+\bq_2}{2}$. Finally, let us assume that the local Berry curvature $\mathcal{B}_\alpha(\bk)$ can be replaced by its average
\be\label{eq:Bbardef}
\overline{\mathcal{B}_\alpha} = \frac{\int_{BZ} d\bk\, \mathcal{B}_\alpha(\bk)}{\int_{BZ} d\bk } = \frac{2\pi C_\alpha}{{A}_{BZ}}
\ee
over the BZ; here ${A}_{BZ} = c_0^2/a^2$ is the area of the BZ, with $a$ the lattice spacing and $c_0$ a numerical constant depending on the unit cell symmetry.
This yields,
\be\label{eq:Cherncomm2}
\left[ \overline{\rho}_{\bq_1},\overline{\rho}_{\bq_2}\right] \approx i \, \bq_1\wedge\bq_2  \overline{\mathcal{B}_\alpha}\,\overline{\rho}_{\bq_1+\bq_2}.
\ee
which is identical to the long-wavelength limit of the density algebra (\ref{eq:GMP}) for the LLL, with  $\overline{\mathcal{B}_\alpha}^{1/2} =\frac{\sqrt{2\pi C_\alpha}}{c_0} a$ playing the role of the magnetic length
$\ell_B$.

Eqn.~(\ref{eq:Cherncomm2}) is our central result and various remarks are in order concerning it:
\begin{enumerate}
  \item{We can define a coarse grained, projected, position operator as $\br_{cg} \equiv x_{cg} \xhat + y_{cg}\yhat
={\displaystyle \lim_{\bq \rightarrow 0}} \frac{\nabla_q}{i} \overline{\rho}_{\bq}$. It follows from Eqn.~(\ref{eq:Cherncomm2}) that
\be
[x_{cg}, y_{cg}] =- i\overline{\mathcal{B}_\alpha}
\ee
which identifies the $\br_{cg}$ with the guiding center position operator in the LLL.}
\item{ For a system of $N$ unit cells there are $N$ points in the BZ. If $\mathcal{B}_\alpha(\bk)$ is truly constant we can define a set of $N$ parallel translation
operators $T_\bq$ for which (\ref{eq:approxparalleltransport}) holds exactly:
\be\label{eq:exactparalleltransport}
T_\bq \ket{\bk, \alpha} = e^{i \int_{\bk}^{\bk+\bq} d\bk' \cdot\mathcal{A_\alpha}(\bk')} \ket{\bk +\bq, \alpha} \ .
\ee
The algebra of the $T_\bq$ is thus exactly of the $\Winf$ form (\ref{eq:GMP}) without a long-wavelength restriction. We note that the $T_\bq$  are trivially isomorphic to magnetic
translation operators for a system with $N$ sites and flux $1/N$ per unit cell and it is straightforward to check that the states in the band form an $N$ dimensional irreducible
representation of their algebra \cite{Brown:1964p1}. From this perspective the idealization of a constant curvature CB hosts a $\Winf$  algebra whose long-wavelength generators coincide
with the physical density operators.}
\item{ Alternatively, if the deviation of the Berry curvature from its average value is bounded, $|\mathcal{B}_\alpha(\bk) -\overline{\mathcal{B}_\alpha}| < |\overline{\mathcal{B}_\alpha}| -\epsilon$, we may define a  `smoothed' density operator which may be regarded as the projection of an operator $\rho^s(\br)$ local in position space; for $qa\ll1$  this gives a modified form of (\ref{eq:approxparalleltransport})
\be\label{eq:smootheddensities}
\overline{\rho}^s_\bq \ket{\bk,\alpha} = \frac{\overline{\mathcal{B}_\alpha}}{{\mathcal{B}_\alpha(\bk)}}e^{i \int_{\bk}^{\bk+\bq} d\bk' \cdot\mathcal{A_\alpha}(\bk')} \ket{\bk +\bq, \alpha}.
\ee
At long wavelengths, the algebra  of smoothed densities closes, and in this limit (\ref{eq:Cherncomm2}) is an {\it exact} equality when $\overline{\rho}_\bq$ is replaced by $\overline{\rho}^s_\bq$. }
\item{ Both the ``Hofstadter'' and ``Haldane'' problems give rise to (\ref{eq:Cherncomm2}), which unifies  earlier lattice FQHE studies \cite{Sorenson:2005p1, Moller:2009p2} with the ones considered here.}
\item{ This last observation can be used to get nearly constant curvature bands by approaching the Landau level limit on the lattice, i.e. by picking flux
$1/q$ per plaquette and working the lowest subband at large $q$. While it is impossible to find a constant-curvature CB in  models of Chern insulators with $\mathcal{N}=2$ bands, it is possible to construct models with $\mathcal{N}>2$ which host a CB with nearly constant Berry curvature \cite{RahulUnpub, RegnaultBernevigUnpub}.}
\item{ Finally, readers familiar with the LLL problem will note that there the $\Winf$ algebra in a system with $N$ states is generated by $N^2$ density operators while in the CB
there are only $N$ densities (or $T_\bq$ if one wishes to work with a closed algebra). This distinction arises as the LLL is formally defined on a continuous space but
is without fundamental dynamical significance as the relevant momenta, $q \ell_B < 1$ are $O(N)$ in the LLL as well. For example, it was shown in \cite{Boldyrev:2003p1} that
keeping only this set of momenta keeps the entire physics of the quantum Hall localization transition in the LLL. However this counting discrepancy does have the consequence
that the algebra of the densities themselves {\it must} close in the LLL at  {\it all} $\bq$ which is not the case in the CB.}
\end{enumerate}

\noindent{\bf Interactions and Disorder:} We have argued above that it is possible to construct lattice models with nearly constant curvature for which the
algebra (\ref{eq:Cherncomm2}) is realized to an excellent approximation. Evidently if we can ignore the variations in curvature and replace the $\overline{\rho}_{\bq}$
by $T_\bq$ in the Hamiltonian (\ref{eq:LLLHamiltonian}), we will have mapped accurately between the LLL and CB problems at the same fractional filling.
Our remaining task is to argue that we can often ignore the residual variations anyway. For a problem with sufficient disorder, most notably the problem of localization
within the CB, we expect that impurity scattering will effectively lead to an averaging of the curvature over the band. This is consistent with what is known about
the QH transition in Chern insulators \cite{Onoda:2003p1} although a direct test in the projected CB would be desirable. A simple estimate for the requisite strength of disorder can be made by comparing the inverse mean free path $l_\text{MF}^{-1}$ to the characteristic momentum-space scale $k_\sigma \sim \overline{|\nabla \mathcal{B}|}/\overline{\mathcal{B}}$ for variations of the Berry curvature; for disorder sufficiently strong that $l_\text{MF}^{-1}\sim k_\sigma$, the random potential will scatter between points in the BZ that are $k_\sigma$ apart, and thereby average their Berry curvature. For the formation of FQH states we appeal to the stability of topologically
ordered states to arbitrary perturbations. Intuitively, we expect to be able to trade the extra pieces in the density commutators for terms in the Hamiltonian. For
sufficiently constant curvature bands and sufficiently strong quantum Hall states, these extra terms should not destabilize the FQHE;  numerical studies are in striking accord with this observation \cite{RegnaultBernevigUnpub, Wu:2011p1}.

\noindent{\bf Concluding Remarks:} This Letter has been concerned with establishing a correspondence between a CB and the LLL; the existence of QH physics in both problems is  traced
to a common source in the nontrivial, $\Winf$, algebra of long-wavelength densities. Intuitively, the nontriviality of the algebra is needed for the densities to do something other than
simply condense and break translational symmetry when an interaction is introduced into a flattened band. This suggests that a program of identifying density algebras for other topologically 
nontrivial bands (sets of bands) in various dimensions could be fruitful in generating new physics upon inclusion of interactions when such bands are partially filled. Here we make a few comments in that direction.

Consider projecting onto a set of bands. For an interesting dynamical problem to arise it will be necessary for these bands to be nearly degenerate (possibly exactly on account of symmetry)
and nearly flat but we are again interested in the kinematics of projection. The projected density operator $\bar{\rho}_{q}$ now has a nontrivial structure in the band index,
so that the analog of (\ref{eq:approxparalleltransport}) is
\be
\overline{\rho}_\bq \ket{\bk,\alpha} \approx \sum_{\beta}\left[e^{i \int_{\bk}^{\bk+\bq} d\bk'\cdot \mathcal{A}(\bk')}\right]_{\beta\alpha} \ket{\beta,\bk+\bq}
\ee
where $\mathcal{A}_{\alpha\beta}(\bk) \equiv  \sum_b u^{\alpha*}_b(\bk)  \frac{\nabla_\bk}{i}u^\beta_b(\bk)$ is the non-abelian vector potential, and the term in square brackets is understood as a matrix exponential. At long wavelengths, 
\be
\left[ \overline{\rho}_{\bq_1},\overline{\rho}_{\bq_2}\right] &\approx&  i \, \bq_1\wedge\bq_2 \sum_{\bk,\alpha,\sigma,\beta} \mathcal{F}_{\alpha\sigma}(\bk) \left[\sum_b u^{\sigma*}_b\left(\bk_+\right)u^{\beta}_b\left(\bk_-\right)\right.\nonumber\\ & &\,\,\,\,\,\,\,\,\,\,\,\,\,\,\,\,\,\,\,\,\,\,\,\,\,\,\,\,\left. \times\gamma^\dagger_{\bk_+,\sigma}\gamma_{\bk_-,\beta}\right]
\ee
where  $\mathcal{F}= d\mathcal{A} - i[\mathcal{A},\mathcal{A}]$ is the non-abelian Berry curvature.  Let us specialize to cases where $\mathcal{F}$ is nearly constant up to
a gauge transformation. It can then be replaced by its BZ average $\overline{\mathcal{F}}$, and the long-wavelength algebra simplifies
to an intertwined generalization of $\Winf$:
\be 
\left[
  \overline{\rho}_{\bq_1},\overline{\rho}_{\bq_2}\right] \approx i \,
\bq_1\wedge\bq_2
\overline{\mathcal{F}}\cdot\overline{\rho}_{\bq_1+\bq_2}.  
\ee
where we have assumed a matrix product in the band indices.

In two dimensions, $\mathcal{A}(k)$ and $\mathcal{F}$ can always be
globally diagonalized by gauge transformation. Hence in the long-wavelength 
limit, the density operators can be labeled by a new ``band'' index, such 
that densities with different indices always commute while those with the same
index exhibit the standard $\Winf$ commutators. If the interactions are chosen
to respect this structure, one obtains a mapping to a multicomponent quantum 
Hall system with different components potentially experiencing different strength
magnetic fields. For example, in the case of topological insulators with
time-reversal symmetry in two dimensions, a basis of time-reversal conjugate bands may always
be found \cite{Roy:2009p1} where the two bands have odd Chern numbers, which are the
same in magnitude and opposite in sign and this leads to the fractional quantum spin Hall 
effect \cite{Levin:2009p1}. We note though that generic interactions will not decompose naturally
in such a basis.
For $d>2$, it is not in general possible to globally diagonalize the Berry connection and now 
the problem is intrinsically non-abelian and worthy of further study. We conjecture that in $d=4$ 
this approach will lead to fractional states for bands with a non-zero second Chern number, possibly 
related to the ones studied in \cite{Zhang:2001p1}.

\noindent{\bf Acknowledgements:} We are grateful to Steve Simon and Duncan Haldane for 
insightful remarks, and to Andrei Bernevig for stimulating discussions and a careful reading of the manuscript. We would like to acknowledge support from the  NSF  through grants DMR-1006608 and PHY-1005429 (SAP, SLS) and the EPSRC through grant 
EP/D050952/1 (RR).

\bibliography{FractionalCI_bib}
\end{document}